\documentclass[twoside,11pt]{article}
\usepackage{a4wide}
\usepackage{epsfig}
\usepackage{pstricks}
\usepackage{pst-node}

\begin{document}
\pagestyle{myheadings}

\markboth{WLPE'01}{Combining Propagation Information and Search
Tree Visualization}

\title {Combining Propagation Information and Search Tree Visualization using ILOG OPL Studio
\footnote{In A. Kusalik (ed), Proceedings of the Eleventh Workshop
on Logic Programming Environments (WLPE'01), December 1, 2001,
Paphos, Cyprus. COmputer Research Repository
(http://www.acm.org/corr/), cs.PL/0111040; whole proceedings: cs.PL/0111042.}}
\author{Christiane Bracchi, Christophe Gefflot, Fr\'{e}d\'{e}ric
Paulin\\
ILOG 9 rue de Verdun 94253 Gentilly Cedex France\\
cbracchi@ilog.fr, cgefflot@ilog.fr, fpaulin@ilog.fr}
\date{}
\maketitle

\begin{abstract}
In this paper we give an overview of the current state of the
graphical features provided by \textsc{ILOG OPL Studio} for
debugging and performance tuning of OPL programs or external
\textsc{ILOG Solver} based applications. This paper focuses on
combining propagation and search information using the Search Tree
view and the Propagation Spy. A new synthetic view is presented:
the Christmas Tree, which combines the Search Tree view with
statistics on the efficiency of the domain reduction and on the
number of the propagation events triggered.
\end{abstract}

\section*{Introduction}
Built upon the \textsc{ILOG Views} graphical components
\cite{ViewsRefMan}, \textsc{ILOG OPL Studio} was initially
intended to provide an Integrated Development Environment for the
OPL language (Optimization Programming Language)
\cite{VanHentenryck99}. Now it can also serve as a debugging
environment for external \textsc{ILOG Solver} applications
independently of the OPL language. Based on new facilities
provided in \textsc{ILOG Solver} 5.0 \cite{SolverRefMan}, such as
the search monitor and the trace mechanism, \textsc{ILOG OPL
Studio} provides tools for graphically visualizing and debugging
the execution of an \textsc{ILOG Solver} program. This paper
explores both OPL program debugging and pure \textsc{ILOG Solver}
application debugging using the \textsc{ILOG OPL Studio} debugger.
However, the graphical features presented could be reused with
other constraint solvers.

In CP programs there is a two-level architecture consisting of a
constraint component and a programming component
\cite{VanHentenryck99}. The constraint component, also called
constraint store, contains the constraints accumulated at a given
computation step and provides basic operations that lead to domain
reductions by means of constraint propagation. The constraint
component is also responsible for maintaining constraint
satisfaction and detecting failures. The programming component
provides a means of combining the basic operations, often in a
non-deterministic way, to guide the solver in the search for a
solution. This two-level architecture leads to a two-level
visualization. The programmed search part can be visualized as a
direct representation of the Search Tree \cite{Aggoun97,Sch97}.
The constraint propagation can lead to a representation of the
impact on the domains of the decision variables
\cite{Aggoun97,Carro99,OPLStudioUserMan} and to a trace of the
events \cite{Ducasse}. This paper focuses on combining propagation
and search information. Although \textsc{ILOG OPL Studio} provides
several graphical facilities for domain visualization, these
features are not presented here. First, we describe the Search
Tree visualization. Next, we show how the Propagation Spy
represents the propagation events. Finally, we show how to combine
search tree and propagation information in terms of both user
interaction and integrated visualization. Special attention is
paid to providing an interaction mechanism familiar to users of
debuggers for traditional languages. A new synthetic view is
presented: the Christmas Tree, which combines statistics on domain
reduction and propagation events with the Search Tree view.

\section*{Visualization of the Search Tree}
Different approaches to search tree visualization are possible,
from dynamic visualization to post-mortem analysis
\cite{Sch97,Aggoun97}. We choose the dynamic visualization
approach so that we can monitor the exploration order. Because the
educational aspect is important in such a tool, and because
\textsc{ILOG Solver} exploration strategy is not limited to
Depth-First Search (DFS) \cite{PerronCP2000}, we should be able to
visualize which is the current visited node, in which order the
tree is explored and in which state each node currently is. We use
colors to indicate the state of the node and a yellow arrow to
indicate the current node. When using exploration strategies other
than the DFS you can see the yellow arrow jumping in the tree. The
possible states of a node are: created but unexplored so far
(white), explored (blue), pruned without exploration (black),
solution (green), failure (red). When the yellow arrow points to
the root, the algorithm is performing the initial domain
reduction.
\begin{figure}
  \centering
    \includegraphics[scale=0.4]{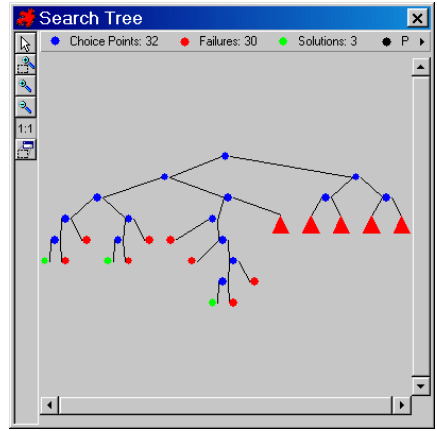}
  \caption{Search Tree visualization for a Golomb Ruler problem with five collapsed subtrees}\label{Figure 1}
\end{figure}
\subsection*{Tree Reduction}
Because the search tree can be very big, \textsc{ILOG OPL Studio}
provides ways of obtaining a simplified version of the tree. For
instance, it is possible to collapse or expand a subtree. A
collapsed subtree is represented by a triangle following the same
color conventions as the node representation (Figure 1). Also, OPL
constructs provide a way to achieve search tree reduction by
abstracting the internal binary tree as an n-ary tree provided
that the search procedure contains the \texttt{forall} and
\texttt{tryall} OPL keywords. \texttt{forall} is an iterated "AND"
and \texttt{tryall} is an iterated "OR", specifying an AND-OR tree
\cite{VanHentenryck2000}. The reduction to the n-ary OR-tree is
not obtained by applying abstraction operations on the initial
binary tree as exposed in \cite{TreeAbstraction} but by means of
interlaced goals keeping the information on the true common father
of the choices. See Figure 2.
\subsection*{The Choice Stack}
Visualizing the shape of the tree is rarely sufficient for program
debugging and performance tuning. One needs to know which choice
occurred at a specific node and the list of choices made so far in
the current branch. This leads to the notion of "Choice Stack". In
\textsc{OPL Studio}, the Choice Stack is a panel describing the
choices made from the root in the current branch. Figure 2 shows
the Choice Stack and the Search Tree for a problem of warehouse
location \cite{SolverRefMan}. Here, the choices are visible in the
form \textit{variable = value}. After double-clicking on a node in
the Search Tree view, the corresponding frames are highlighted in
the Choice Stack panel. In Figure 2, the first node below the root
is selected and the corresponding frames in the Choice Stack show
that the choice at the selected node is the value \texttt{London}
set to the variable \texttt{Supplier[0]}. The depth from the root
is given between brackets.
\begin{figure}
  \centering
    \includegraphics[scale = 0.35]{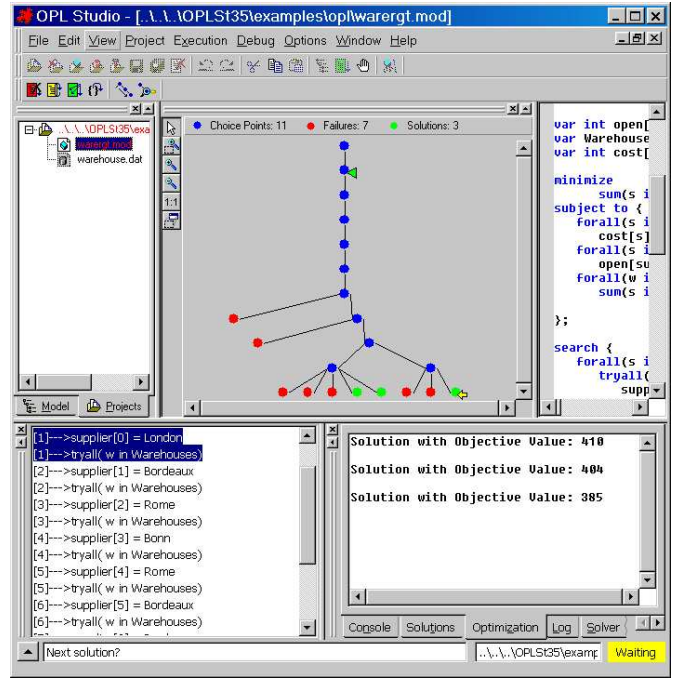}
  \caption{Choice Stack and n-ary Search Tree for a warehouse location
  problem
 }\label{Figure 2}
\end{figure}
\section*{Visualization of Propagation Events}
Constraint propagation is called before the search and at each
Search Tree node. The propagation occurring before the search, at
the root node, is called the \textit{initial propagation}. The
\textsc{ILOG OPL Studio} debugger provides a way of displaying the
trace of the propagation events and the result of the propagation
in terms of domain reduction: the Propagation Spy.
\subsection*{The Propagation Spy}
Based on the new Trace mechanism provided in \textsc{ILOG Solver}
5.0 \cite{SolverRefMan}, the \textsc{ILOG OPL Studio} debugger
provides a panel called the "Propagation Spy". This is a special
hierarchical sheet containing a tree hierarchy in the first column
\cite{ViewsRefMan}. Figure 3 shows the initial propagation for the
"Pheasants and Rabbits" problem \cite{CHARME}. Each line is a
Propagation Event, such as "Set Max", "Set Value", "Post
Constraint", "Propagate Constraint", "Constraint Fail", etc. Each
column contains the impact on one variable. Time passes
vertically, from top to bottom. The textual description of the
events is appended inside the first column. When an event line is
selected the columns are rearranged so that the variables of
interest are displayed first. The cell at the junction of an event
line and a variable column is colored according to the type of
event and shows the result of the event in textual form. Thus, the
domain reduction history can be followed here.
\begin{figure}
  \centering
    \includegraphics[scale = 0.5]{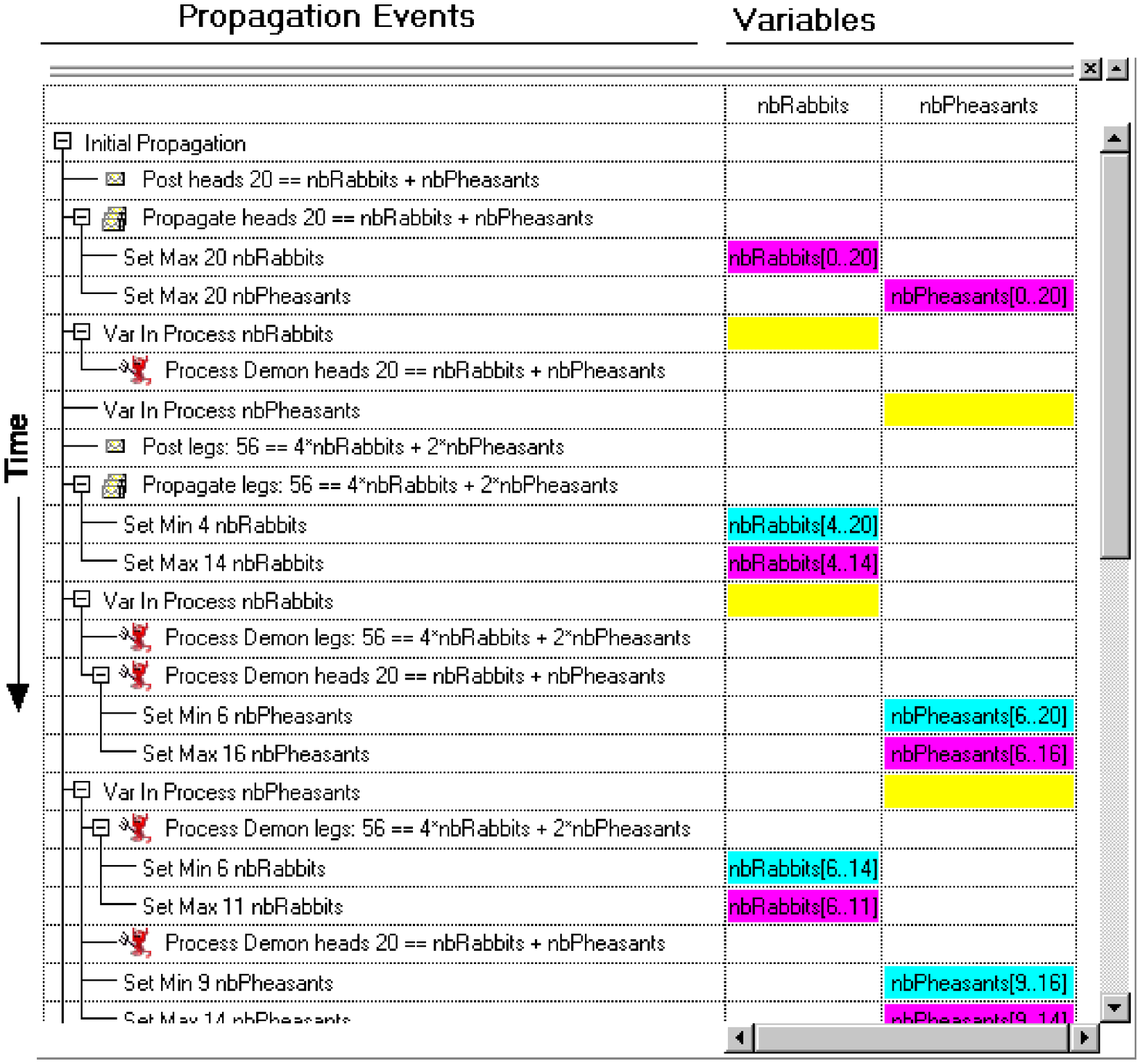}
  \caption{The Propagation Spy for the pheasants and rabbits problem}\label{Figure 3}
\end{figure}
Because the number of events can be very high, the Propagation Spy
is not always active: it is activated on demand by the user. We
describe the user interaction in more detail in the next section:
\textit{Combining Search and Propagation Information}. The
Propagation Spy is convenient for propagation teaching and for
failure analysis. When used with \textsc{ILOG Solver} add-ons such
as \textsc{ILOG Scheduler}, a specific trace is inserted in the
Propagation Spy \cite{SchedulerRefMan} so that the program can be
debugged at a higher level.
\section*{Combining Search and Propagation Information}
In this section we will show how the user interacts with the
debugger with the two levels of information: the search and the
propagation. Then we will propose a synthetic view combining
statistical information about the propagation with the search
tree.
\subsection*{User interaction with the debugger}
We took special care to provide an interaction mechanism familiar
to users of debuggers for traditional languages (such as Microsoft
Visual Studio, Karmira BugSeeker$^\texttt{TM}$ for Java or dbx on
UNIX). In such debuggers, the user can interrupt the execution,
step into, out of, or over a function, and restart. In debuggers
for traditional languages the function call stack is a common
abstraction of the flow of execution, and the current state of the
variables can be inspected. Here, in the context of a CP program,
these concepts are respectively replaced by the choice stack and
the state of the domains. In Figure 4 we propose a specification
for user interaction with a CP debugger such as the \textsc{ILOG
OPL Studio} debugger and a mapping to traditional debugging
concepts in a traditional debugger.
\begin{figure}
  \centering
  \footnotesize
\begin{tabular}{|c|c|c|}
  \hline
  \textbf{Action} & \textbf{Traditional debugger} & \textbf{CP debugger} \\
  \hline
  Step Into & Step into a function & Step into a node, tracing prop. \\
  \hline
  Step Out & Step out of a function & Step out of a node, stop tracing prop. \\
  \hline
  Step Over & Step over a function & Step over a node, or prop. event  \\
  \hline
  Skip Step & Step skipping predefined files & Trace prop. and stop at next node \\
  \hline
  Break  & Break at next instruction & Break at next node or event \\
  \hline
  Breakpoint & Break at line & Break at node \\
  \hline
  Current & Current line in the source code & Current visited node \\
  \hline
  Stack & Function Call Stack & Choice Stack \\
  \hline
\end{tabular}
  \caption{Concept mapping between a traditional debugger and a CP debugger}\label{Figure 4}
\end{figure}
When the user launches a first execution, he sees the tree drawing
itself in the Search Tree viewer. If necessary, he can interrupt
the execution. The debugger stops at the beginning of a node
visit. Then, if the user chooses to step into the node, each step
is a propagation event. So a line is added to the Propagation Spy
at each "Step Into" command. He can then step out, that is, stop
tracing the propagation and continue to the next visited node.
Because of the non-deterministic nature of the CP search, it is
not possible to draw the tree in advance. A second run is
necessary to focus on a segment of the tree. After a first run,
the user identifies interesting regions and places a breakpoint at
an interesting node. At the second run, the tree nodes are
"laundered", (i.e. they become white) and then recolored. The
debugger stops at the breakpoint, that is, when the corresponding
Search Tree node is visited.
\subsection*{A synthetic view: the Christmas Tree}
The Search Tree viewer of \textsc{ILOG OPL Studio} provides a way
of combining statistics on propagation information with the tree
representation. The information added to the tree is the number of
propagation events and the effective global domain reduction at
each node. It is important to distinguish between these two pieces
of information because many propagation events can be triggered
with little impact on the domain reduction. The size of the nodes
becomes proportional to the number of propagation events fired at
each node (which is highly correlated with the time spent at each
node). The meaning of the colors remains unchanged, except that
the color is lighter or darker depending on the effective domain
reduction obtained during the propagation at this node. The Search
Tree now has big and small balls with different colors, that is
why we call it the "Christmas Tree". Figures 11 and 12 represent
the Christmas Trees associated with the "Golomb Ruler" problem
\cite{golomb82shift}. The statistical information connected with
the initial propagation is concentrated at the tree root. It
becomes obvious that all nodes are not equal. We can detect if, in
failure nodes (red nodes), the failure is discovered early (small
node) or late (large node).
\section*{Principle of Operations}
In this section we describe the basic implementation structure of
the visual tools and their integration with user applications.

The architecture is a client-server architecture where the GUI is
the server and the user application the client. The GUI is loosely
coupled with the application. It could be reused with non-ILOG
products, provided that the protocol between the GUI and the
engine is respected. This protocol is based on XML messages
exchanged via sockets. For ILOG based applications, this protocol
is encapsulated in a C++ library called "Debugger Library" which
takes care of the communication layer and must be linked to the
user application. The API is object-oriented, as is \textsc{ILOG
Solver} \cite{puget95beyond}. The application programmer must
implement an \texttt{IlcDebuggable} interface, overriding two
virtual methods, \texttt{stateModel()} and \texttt{solveModel()}.
He must instantiate an \texttt{IlcDebugger} object passing as
arguments the name of the machine and the port number on which the
GUI server is listening. The Debugger Library has its own
subclasses of the Search Monitor class and of the Propagation
Trace class defined in \textsc{ILOG Solver} \cite{SolverRefMan}.
These classes have a set of virtual methods for each search and
propagation event. The Debugger classes take care of sending the
appropriate Propagation and Search information to the GUI,
wrapping it in XML messages.
\section*{Experimentation}
In this section, we will concentrate first on search procedure
improvement with the regular Search Tree, using a scheduling
problem expressed in OPL. Then, we will show how to use the
Christmas Tree and the Propagation Spy to illustrate propagation
issues with a pure \textsc{ILOG Solver} sample.
\subsection*{Improving the search} The
experimentation for search improvement is based on a Scheduling
problem: job-shop 6 \cite{Fisher63}. The aim is to schedule a
number of jobs on a set of machines to minimize completion time,
often called the \texttt{makespan}. Each job is a sequence of
tasks and each task requires a machine. Each intermediate solution
found by OPL improves the objective function (\texttt{minimize
makespan}) while satisfying the other constraints. Figure 6 shows
the shape of the tree with the default search procedure of OPL. To
begin with, we basically have the intuition that we must pay
attention to \textbf{right subtrees} (Figure 5).
\begin{figure}
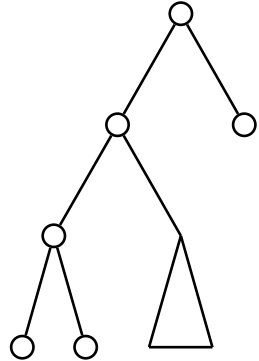

\psset{linewidth=1pt}
\def\CN#1{\cnode{1ex}{#1}}
\begin{center}
\begin{tabular}{cccccccc}
&&&&&\CN{n1} \\ [1cm]
&&& \CN{n2} &&&&  \CN{n3} \\ [1cm]
&\CN{n4} &&&&  \pnode{n5} \\ [1cm]
\CN{n6} && \CN{n7} && \pnode{n8} && \pnode{n9}
\end{tabular}
\ncline{-}{n1}{n2}
\ncline{-}{n1}{n3}
\ncline{-}{n2}{n4}
\ncline{-}{n2}{n5}
\ncline{-}{n4}{n6}
\ncline{-}{n4}{n7}
\ncline{-}{n5}{n8}
\ncline{-}{n5}{n9}
\ncline{-}{n8}{n9}
\end{center}
\caption{Right sub-tree in a binary tree}\label{Figure 5}
\end{figure}
For a binary tree, if we consider that the right branch is the
contradiction of the left branch, when right subtrees are
developed one of the following two conclusions can be drawn.
Either solutions are found in the right subtree, in which case the
search could be improved since it is the contradiction of a choice
that leads to a better solution. Or, solutions are not found in
the right subtree, which is an indication that the propagation
could be improved, because the back propagation of the cost should
have pruned the subtree automatically. Therefore, the presence of
right subtrees should be monitored. Next, when using the Depth
First Search exploration order, the higher in the tree the right
subtrees are, the more we should pay attention to them, because
the tree is potentially exponential in its depth.
\begin{figure}
  \centering
    \includegraphics[width=3.62in,height=1.70in]{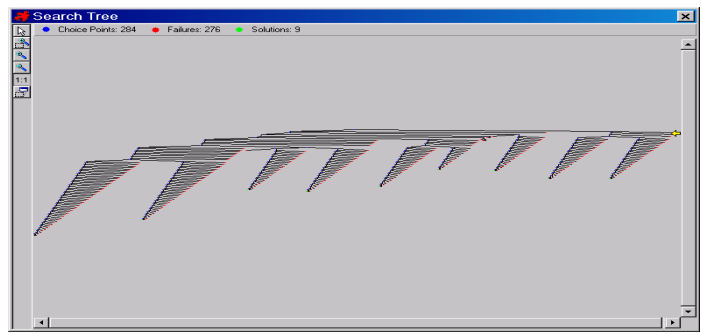}
  \caption{The job-shop 6 problem search tree using the default search procedure of OPL}\label{Figure 6}
\end{figure}

When we use the default search procedure of \textsc{ILOG OPL
Studio} to
 solve the job-shop 6 problem (with Depth First Search as the exploration order), we notice that right subtrees exist in
 the higher part of the tree (Figure 6). These right subtrees
 contain solutions, therefore contradictions of the choices made by the default strategy of OPL lead to
 better solutions during the minimization process. In scheduling applications with unary resources, the general strategy
 is to rank each unary resource. Ranking a unary resource consists of finding a total ordering for all activities
 requiring the resource. Once activities are ordered, a solution can be found efficiently. The detected inefficiency of the
 default search procedure is due to the fact that it ranks a resource completely before selecting another resource to rank.
 The default search procedure is not opportunistic enough. The user-defined search procedure in Figure 7 tries to improve
 that.
\begin{figure}
\footnotesize
\begin{verbatim}
search {
  while not isRanked(tool) do
    select(r in 1..6 : not isRanked(tool[r])
                     ordered by increasing nbPossibleFirst(tool[r]))
       select(t in toolTasks[r]
                     : isPossibleFirst(tool[r],task[t.i,t.j])
                     ordered by decreasing dmin(task[t.i,t.j].start))
          tryRankFirst(tool[r],task[t.i,t.j]);
  forall(i in 1..6)
    forall(j in 1..6)
       task[i,j].start = dmin(task[i,j].start);
};
\end{verbatim}
  \caption{A user-defined search procedure for the job-shop 6 problem}\label{Figure 7}
\end{figure}
First, we select the resource that has the smallest number of
activities we could rank at first position. Then, we select among
the tasks requiring this resource that are rankable first, the one
which has the earliest starting time. The instruction
\texttt{tryRankFirst(tool[r],task[t.i,t.j])} is nondeterministic
and has two alternatives. The first alternative adds the
constraint that \texttt{task[t.i,t.j]} be ranked first among the
unranked activities of the \texttt{tool[r]} resource. The second
alternative (used when backtracking) adds the constraint that
\texttt{task[t.i,t.j]} must not be ranked first among the unranked
activities of \texttt{tool[r]}. We made a deliberate mistake in
the task selection by ordering by latest starting time
(\textbf{decreasing} keyword) instead of earliest starting time
(\textbf{increasing} keyword). This leads to the Search Tree of
Figure 8. The tree is worse in shape, with a lot of right
subtrees.
\begin{figure}
  \centering
    \includegraphics[scale=0.3]{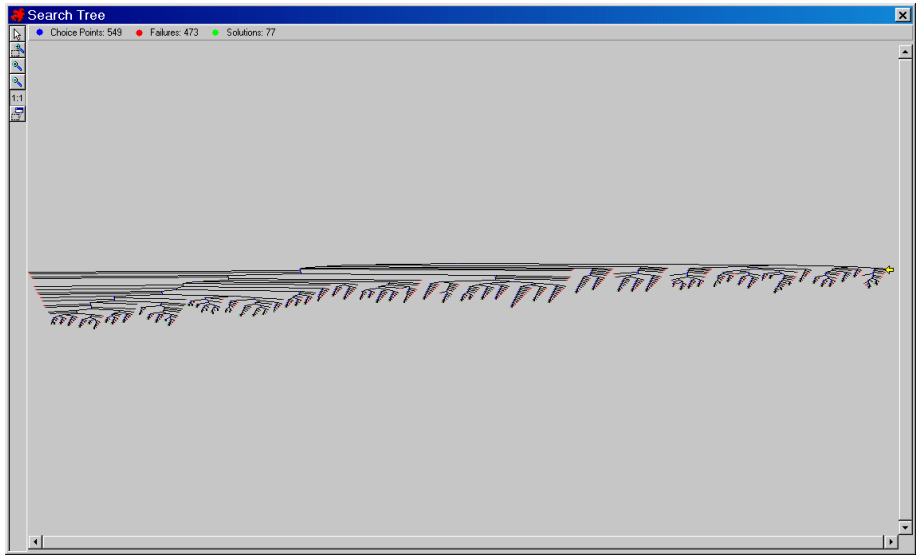}
  \caption{The job-shop 6 problem search tree using a (buggy) user-defined search procedure in OPL}\label{Figure 8}
\end{figure}
Using the Choice Stack panel, we can identify which choice leads
to right subtrees. Then, after ordering by the increasing earliest
starting time, the shape of the tree is much better. There are
fewer right subtrees and the overall size of the tree is smaller
(Figure 9).
\begin{figure}
  \centering
    \includegraphics[scale=0.5]{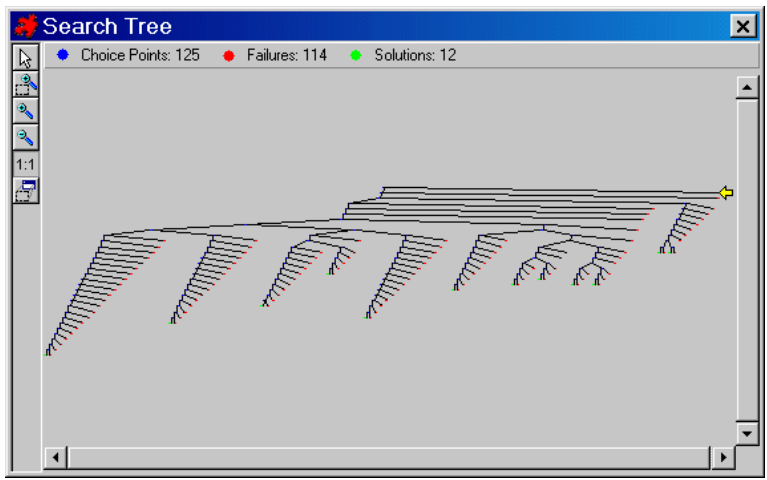}
  \caption{The job-shop 6 problem search tree using an improved user-defined search procedure in OPL}\label{Figure 9}
\end{figure}

\subsection*{Improving the propagation}
The experimentation concerning the propagation improvement is
based on the Golomb Ruler problem \cite{golomb82shift}. The goal
is to find a set of values representing the graduations of a rule
such that the difference between each pair of graduations is
always distinct, and such that the length of the rule is minimal.
We use the \texttt{alldifferent} global constraint of \textsc{ILOG
Solver} with two different levels of propagation. When using the
basic filter level of this constraint, the solver guarantees that,
at any computation point, the specified variables do not have the
values of the already-assigned variables inside their domain. When
using the extended filter level, the solver reasons on the domains
instead of the values and guarantees that, for each value in the
domain of any given variable, there exist values in the domains of
the remaining variables such that the constraint is satisfied. So
the extended level enforces a stronger pruning than the basic one.
\begin{figure}
  \centering
    \includegraphics[scale = 0.60]{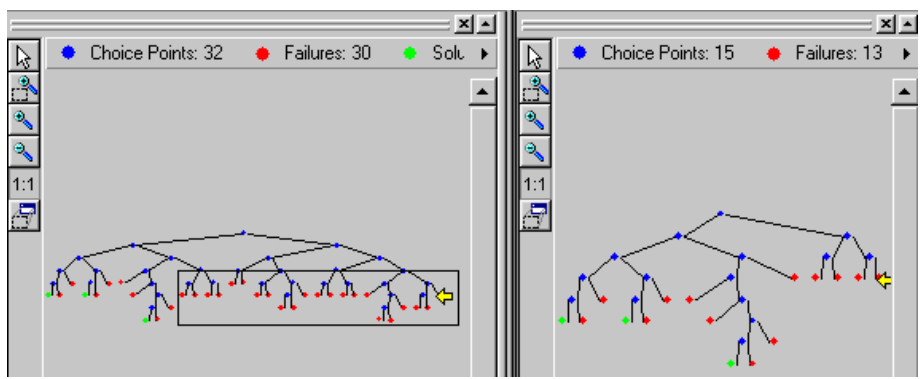}
  \caption{The Golomb problem: on the left the basic filter level,
   on the right the extended filter level}\label{Figure 10}
\end{figure}
Figure 10 represents the trees corresponding to the two filter
levels. The basic filter level produces a bigger tree with
additional right subtrees as compared to the extended filter
level. These right subtrees have only failure leaves, a sign of
lack of propagation. By just setting the filter level to the
extended level these right subtrees are pruned. But does this mean
that we saved time ? Here, the Christmas Tree gives us a picture
of the cost and of the efficiency of this extended propagation.
\begin{figure}
  \centering
    \includegraphics[scale = 0.50]{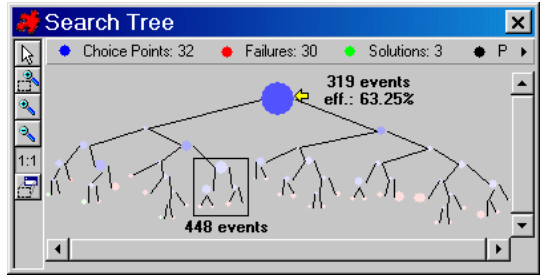}
  \caption{Christmas Tree for the Golomb 6 problem with basic level
   of propagation for the \texttt{alldifferent} constraint}\label{Figure 11}
\end{figure}
Let us compare the number of propagation events in the first right
subtree located in the frame in Figure 11 with the number of
propagation events occurring at the corresponding big failure node
in Figure 12. We see that the big node required two times fewer
propagation events for detecting failure than the subtree. So,
yes, the extended propagation saved time here. Yet, if we compare
the initial propagation statistics by looking at the root node, we
see that the efficiency of domain reduction is the same (63.25\%).
The extended propagation level triggered a few more events but
with no result.
\begin{figure}
  \centering
    \includegraphics[scale = 0.40]{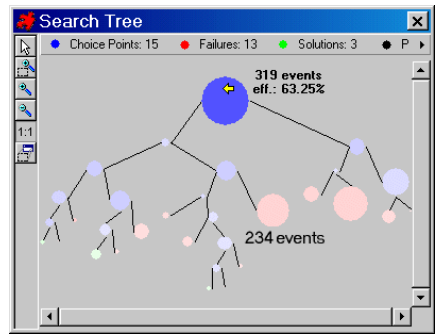}
  \caption{Christmas Tree for the Golomb 6 problem with extended level
   of propagation for the \texttt{alldifferent} constraint}\label{Figure 12}
\end{figure}

Now, let us enter into the details by using the Propagation Spy.
By inspecting the Initial Propagation, we can see that the solver
adds a hidden constraint when posting the \texttt{alldifferent}
constraint. When tracing the propagation at the first big failure
node of the extended filter level and comparing to the
corresponding node of the basic filter level the Propagation Spy
detects the extra propagation (Figure 13). On the left, with the
basic filter level, the propagation stops  after reducing the
variable \texttt{difference[11]} to the interval \texttt{7..11}.
On the right, with the extended level, the propagation of an
additional internal constraint posted by the \texttt{allDifferent}
constraint strongly reduces the domains by means of "Set Min"
events. This additional propagation leads to a failure, avoiding a
subtree exploration.
\begin{figure}
  \centering
    \includegraphics[scale = 0.50]{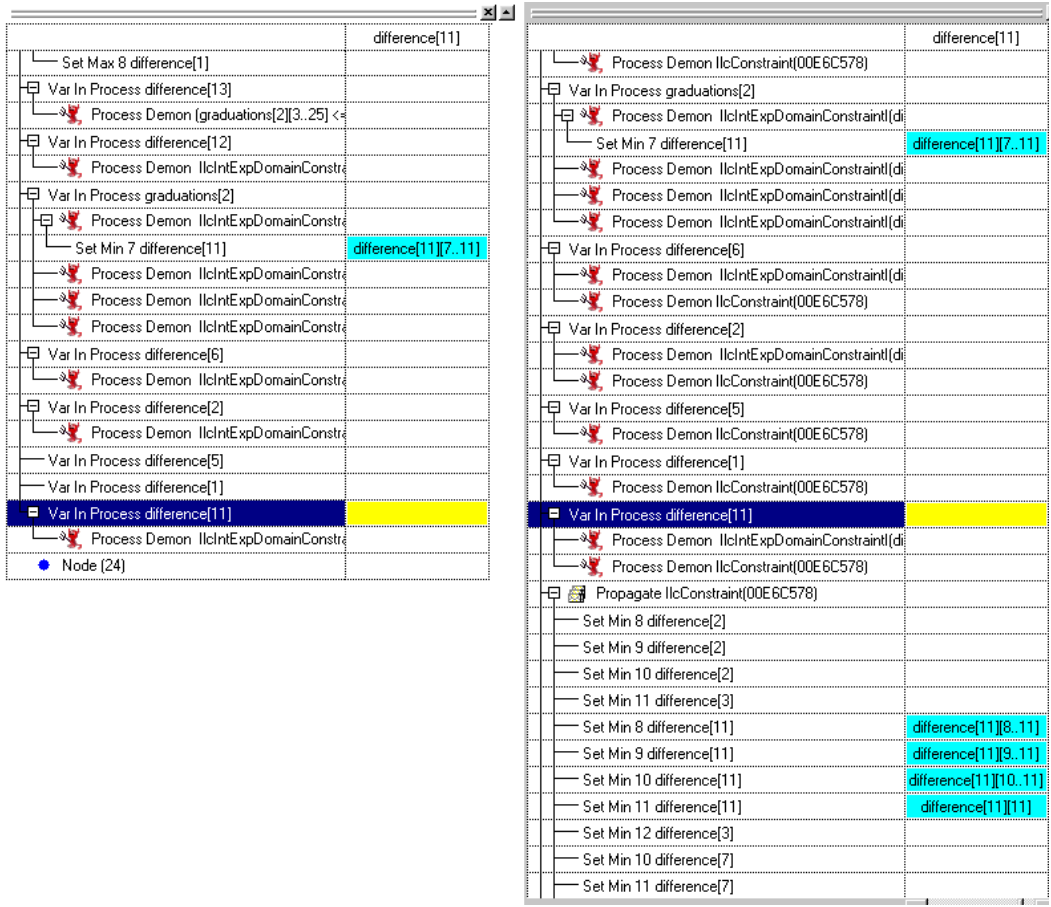}
  \caption{Comparing two Propagation traces using the Propagation Spy for the Golomb 6
  problem (on the left, basic filter level, on the right, extended filter
  level)
}\label{Figure 13}
\end{figure}
However, by inspecting in detail we see that four "Set Min" events
are triggered on the \texttt{difference[11]} variable where one
should suffice. These "Set Min" events are "Remove Value" orders
that have been translated by the solver into "Set Min" because the
value to remove was a bound. Then, we have the intuition that
reasoning on bounds instead of on domains should be sufficient. If
we tune the \texttt{alldifferent} constraint to the intermediate
filter level, which reasons on bounds instead of on domains, we
obtain the same search tree as compared to the extended filter
level. The Propagation Spy shows that the \texttt{difference[11]}
variable at the same Choice Point is bound more quickly (Figure
14).
\begin{figure}
  \centering
    \includegraphics[scale = 1.00]{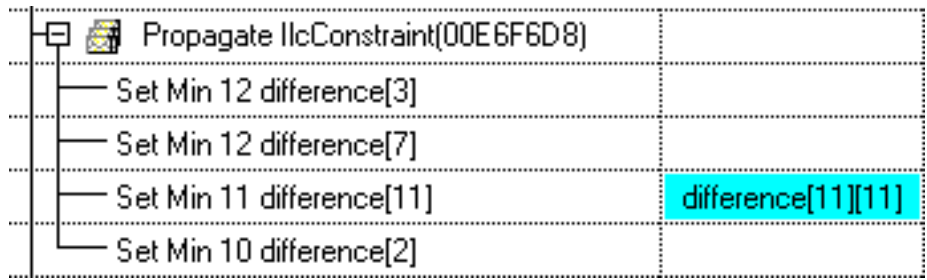}
  \caption{Propagation Spy for the Golomb 6 problem with
  intermediate filter level
}\label{Figure 14}
\end{figure}
\section*{Conclusion}
We have presented an overview of the current state of the
\textsc{ILOG OPL Studio} debugger capabilities for performance
tuning of CP programs. The experimentation on improving the search
procedure with the Search Tree visualization is promising. A
graphic trace of the propagation events, in order to better
understand the inner details of what happens at each node, has
also been presented. Combining the two pieces of information by
adding statistics on the "weight" of each node in the Christmas
Tree seems to be an important step toward visualizing where the
time is spent.
\subsection*{Acknowledgments}
We would like to thank Laurent Perron for his help on the
\textsc{ILOG Solver} search monitor and Jean-Charles R\'{e}gin and
Xavier Nodet for their explanations of the \textsc{ILOG Solver}
and \textsc{ILOG Scheduler} trace mechanism. We would also like to
thank Michel Leconte, Philippe Laborie and Veronica Murphy for
their comments and ideas.

\end{document}